\documentclass[openacc]{rsproca_new}%%%%where rsproca is the template name
\usepackage{comment}
\usepackage[normalem]{ulem} %To strikethrough text \sout{}

\newcommand{\oriol}[1]{{\color{blue}\textbf{#1}}}

\graphicspath{{Figs/}{Logos/}}

\begin{document}

%\title{Past, present and future of emergence in natural sciences. A mini-review.}
\title{From the origin of life to pandemics: Emergent phenomena in complex systems}

\author{Oriol Artime$^{1}$ and Manlio De Domenico$^{2}$}
\address{$^{1}$Fondazione Bruno Kessler, Via Sommarive 18, 38123 Povo (TN), Italy\\
$^{2}$Department of Physics and Astronomy ``Galileo Galilei'', University of Padua, Italy}

\subject{\oriol{TBCompleted}}

%%%% Insert corresponding author and its email address}
\corres{Oriol Artime\\
\email{oartime@fbk.eu};\\
Manlio De Domenico\\
\email{manlio.dedomenico@unipd.it}}

%%%% Abstract text to be placed here %%%%%%%%%%%%
\begin{abstract}
When a large number of similar entities interact among each other and with their environment at a low scale, unexpected outcomes at higher spatio-temporal scales might spontaneously arise. This nontrivial phenomenon, known as emergence, characterizes a broad range of distinct complex systems -- from physical to biological and social ones -- and is often related to collective behavior. It is ubiquitous, from non-living entities such as oscillators that under specific conditions synchronize, to living ones, such as birds flocking or fish schooling. Despite the ample phenomenological evidence of the existence of systems' emergent properties, central theoretical questions to the study of emergence remain still unanswered, such as the lack of a widely accepted, rigorous definition of the phenomenon or the identification of the essential physical conditions that favour emergence. We offer here a general overview of the phenomenon of emergence and sketch current and future challenges on the topic. Our short review also serves as an introduction to the Theme Issue \textit{Emergent phenomena in complex physical and socio-technical systems: from cells to societies}, where we provide a synthesis of the contents tackled in the Issue and outline how they relate to these challenges, spanning from current advances in our understanding on the origin of life to the large-scale propagation of infectious diseases.
\end{abstract}
%%%%%%%%%%%%%%%%%%%%%%%%%%%

%%%%%%%%%%%%%%% End of first page %%%%%%%%%%%%%%%%%%%%%

\maketitle

\section{Introduction}

``\emph{d oijw ao o fyrg bafjdsdpw dweoda wdhao jrfgb sag wdgy d ias dsih sig qqpdjwe fjrfb dvvs}''. In the previous sentence each character is randomly generated: the way characters cluster together, while being separated by spaces, are usually interpreted as words and sequences of words are used to transmit information, e.g. a message. However, there is apparently no intelligible knowledge that can be extracted from the above example: from the perspective of the reader there is a lack of those familiar patterns that one routinely uses to communicate and expects to find with respect to some prior (e.g., a scientific paper written in English). Concisely, we can argue that there is no concept or knowledge in a single character: a sequence of characters (i.e., a word) and a sequence of words (i.e., a sentence) exchanged between a sender and a receiver becomes meaningful when both of them spontaneously start to use repeating patterns which they identify as meaningful. In other words, language is an emergent phenomenon requiring a symbolic representation for units (i.e., the characters) and their interactions (i.e., special sequences of characters).

Rather intriguingly, such a \emph{spontaneous appearance} of meaningful structures in space and time is an ubiquitous phenomenon observed from the microscopic scale -- e.g., in molecular interactions within a cell -- to the macroscopic one -- e.g., the cosmic web, in complex adaptive matter~\cite{cox2005complex}. In the following, we will briefly review the phenomenology concerning emergent phenomena and provide an operational definition of emergence as a hallmark of complexity that can be applied to a variety of complex systems, regardless if they are natural, social or artificial.

\section{A brief historical overview}  

A primordial concept can be already found in the ``Metaphysics'' written by the ancient Greek philosopher Aristotle, where it is argued that a totality is something besides the parts. The same concept, but with a slightly different meaning, can be found in Gestalt psychology, based on the intuition that organisms do not merely perceive individual components but entire patterns or configurations: in a nutshell, ``the whole is other than the sum of the parts''~\cite{koffka1935gestalt}, and a similar concept appears in the work by William M. Wheeler in 1926~\cite{wheeler1926emergent,wheeler1927emergent}. The concept has been invoked by emergentist philosophers such as Samuel Alexander and Charlie Dunbar Broad in contrast to reductionism in the '20s, although a more formal approach was developed by the pioneers of cybernetics in the '40s. In fact, cybernetics deals with systems and their causal feedback loops: among its founders there were Norbert Wiener and John von Neumann, the latter being the first to propose cellular automata and a universal constructor, both concepts strictly related to emergence. Ludwig von Bertalanffy, in particular, was among the founders of general systems theory, providing the first mathematical ground to describe the complexity observed in biological and social systems~\cite{von1950outline,ludwig1951general}.
 
It was during the '70s that the Nobel Laureate Phil Anderson warned against the perils of reductionism. In~\cite{anderson1972more}, he gave specific examples where reductionist thinking fails and highlighted the fact that the most fundamental physical laws were unable to explain new properties and behaviors arising in the assembly of a large number of units obeying those fundamental laws. 
An important consequence of this is that he opened the door to the existence of fundamental laws at different levels of complexity, e.g., the objects of study in biology do not follow the laws of chemistry, likewise the objects of study in chemistry do not follow the laws of particle physics. Anderson argued that, even if we are looking at a single level of complexity in this hierarchy, it is via a process of \textit{symmetry breaking} that the state of a large system composed by many entities might not follow the rules of the fundamental laws that the entities themselves follow. Hence, the appearance of new properties is intimately linked with the disappearance of the symmetries of a system, be them spatial, temporal, informational, etc. As a particular example of emergence by symmetry breaking, we can mention the formation of complex spatio-temporal patterns in dissipative systems, where the isotropic symmetry that one would expect from thermodynamics is broken~\cite{nicolis1977self}. Noteworthy, the development of this theory was contemporary to Anderson's seminal paper, and earned Ilya Prigogine the Nobel prize in 1977. After Anderson's illuminating article, we have witnessed an upsurge of contributions that kept exploring the implications of this concept, both at the theoretical and applied levels, see, e.g.,~\cite{bedau2008emergence} for a compilation of some of these.

One of the directions that Anderson pointed out as interesting to explore is the emergence in living beings. The origin of life can be seen as an instance of enormous complexity whose inception is based on the interaction of cells that perform simple, decentralized tasks. One way to approach this is via computer simulations. Indeed, during the '80s, computers became widespread and thus propelled the exploration of  emergent phenomena from a computational point of view. It is inevitable to mention here the influential investigations of Stephen Wolfram in cellular automata. He introduced the numbering scheme still used nowadays~\cite{wolfram1983statistical} and, among others breakthroughs, he conjectured in 1985 that \emph{Rule 110 cellular automaton} was Turing-complete --- formally proven almost two decades later~\cite{cook2004universality}. In the last years, the research in emergence of life has kept advancing and have incorporated more and more biological and molecular mechanisms.

Last decades have been characterized by the realization that many central problems in Physics, but also in other branches of sciences, could be understood as emergent phenomena, such as superfluidity or the fractional quantum Hall effect~\cite{laughlin1999nobel} who has led Robert Laughlin to be awarded the Nobel Prize for Physics in 1998. The study of emergent phenomena has been made even more popular by Murray Gell-Mann (Nobel Prize for Physics in 1969)~\cite{gellmann1994the} and has turned more and more an interdisciplinary endeavour and have found in the wide umbrella of complexity science a substrate to develop~\cite{de2019complexity}. Nowadays, efforts go in the directions of identifying, characterizing and understanding such phenomena, with a balanced combination of analytical, computational and experimental techniques, as well as providing a formal theory of emergence, with considerable developments made thanks to information-theoretic tools. The last great news for the field, highlighting it is far from being a fringe theory, concerns the Nobel Prize for Physics awarded to Giorgio Parisi in 2021, for his studies of disordered physical complex systems and their fluctuations.

\section{What is emergence and why does it matter?} 

Simple systems are mostly characterized by the fact that the properties of the whole can be understood, deduced or predicted from the analysis of their components in isolation, their addition or their aggregation: in practice, macroscopic observables can be deduced from microscopic ones. From this observation, it is clear that in order to characterize an emergent phenomenon one needs at least two well separated scales -- for instance defined in terms of energy or in space and time -- and one external observer able to identify meaningful patterns, and measure them in terms of information, appearing at one scale but not at the others. Let us consider, for instance, the mass $M$ of composite objects like a chair, which consists of distinct parts with a mass $m_i$ ($i=1,2,...,n$): the overall mass can be simply obtained by summing up the mass of each component as $M=\sum\limits_{i}m_i$. At a smaller scale, let us say at molecular one, a similar approach leads to a similar result. At the lowest scale, like the one of atomic nuclei, one could argue that the same approach would still lead to a similar result, although this is not exactly the case because the strong interaction which combine protons and neutrons -- i.e., the nuclear force -- is responsible for a mass defect which is converted into binding energy according to the mass-energy equivalence. The mass is an interesting property, since at spatial scales much larger than atomic one the linear approximation applies very well, while at the lowest scale it does not. At a fundamental level, like in quantum field theory, mass allows to measure the coupling of a particle with the Higgs field but it is not considered an emergent property, although the issue has been debated~\cite{wilczek2012origins}. It is also interesting that at the nuclear scale, the presence of interactions between fundamental constituents is able to generate a deviation from the na\"{\i}ve expectation that a simple summation applies. It is worth remarking that the mass property can be defined at the level of a single particle as well as at the level of an aggregation of particles, regardless if they are interacting or not. This is also the case for other physical properties such as the spin, for instance. 

However, there are properties that cannot be defined at the level of a single unit, being it a particle, a cell or an individual: such properties are meaningful only at some scale larger than the one defining a single unit. When it is the case, the corresponding phenomena are usually referred to as \emph{emergent}: emergence is considered a fundamental feature of complex adaptive matter, transcending the traditional frontiers of theoretical physics and becoming a landmark in a broad spectrum of disciplines, from biology to neuroscience, from system ecology to economics. In the following, we will briefly review a broad class of complex systems across a variety of disciplines, starting from the quantum realm and then moving to non-quantum systems, including physics, biology, ecology, social and urban sciences. A special focus will be given to results obtained from network science, where several emergent properties are related to non-trivial structure, non-trivial dynamics or their interplay.

\paragraph{Emergence in quantum physical systems.} 
Quantum mechanics is responsible for many fascinating emergent phenomena, such as localization and superconductivity. Regarding the former, it was during the '50s that Phil Anderson suggested that in a sufficiently large lattice, a sufficient amount of disorder prevents standard diffusion of waves~\cite{anderson1958absence}, which is a setup can be effectively realized by means of impurities or defects in semiconductors. As per superconductivity, we know that a superconductor is a material where the collective behavior of particles spontaneously emerge at a characteristic critical temperature: below such a temperature, the material does not exhibit electrical resistance, making these materials suitable for dissipation-free applications. Although some properties are material dependent, all superconductors break the $U(1)$-gauge symmetry down to $\mathbb{Z}_2$ leading to universal properties such as the Meissner-Ochsenfeld effect and off-diagonal long range order. In condensed matter physics, the origin of this phenomenon can be explained by the theory proposed by Bardeen, Cooper and Schrieffer, arguing that pairs of fermions, such as electrons, condensate into strongly interacting particles in the same ground quantum state -- known as Cooper pairs -- at low temperatures~\cite{bardeen1957microscopic}. We refer the interested reader to a recent collection about emergent superconductivity~\cite{SpecialIssue2020}.

At a larger scale, let us consider the case of two or more superconductors placed close enough to each other to be weakly coupled. The behavior of the overall system was unexpected in the '60s: known as Josephson effect, the production of a supercurrent in absence of voltage, was first predicted by Brian Josephson in 1962 and later observed experimentally~\cite{josephson1974discovery}. Such a phenomenon cannot be deduced from the knowledge of each superconductor in isolation: only the presence of weak coupling allows for the spontaneous appearance of such a collective behavior. The quantum Hall effect, i.e., the quantized version of the Hall effect observed in systems at low temperatures is another emergent phenomenon due to collective behavior~\cite{Klitzing1980,yennie1987integral,laughlin1999nobel}, together with the fractional quantum Hall effect~\cite{laughlin1981quantized,tsui1982two,laughlin1983anomalous}. 

\paragraph{Emergence in classical physical, non-living, systems.} Since the pioneering work of Phil Anderson -- using the mechanism of symmetry breaking to argue against reductionist approaches -- and Prigogine on dissipative structures, a plethora of studies provided convincing evidence for the existence of physical systems characterized by the spontaneous appearance of properties that cannot be understood, or predicted, from the full knowledge of system's constituents.

At a classical scale, an emblematic example of emergent phenomenon is the turbulence observed in fluids. For instance, in Rayleigh-B\'enard convection a fluid is heated from below on a planar horizontal surface, leading to formation of metastable convection cells -- known as B\'enard cells -- which spontaneously break rotational symmetry and self-organize into regular patterns~\cite{ahlers2009heat}. Turbulence cannot be defined at the scale of a single fluid unit, and it emerges in a wide spectrum of hydrodynamic and non-hydrodynamic systems, ranging from the Earth's magnetic field to chemical reactions~\cite{swinney1978hydrodynamic}. Remarkably, fully developed turbulence can be reliably described by assuming that the underlying fluctuations cannot be described by a unique scaling exponent but they require a continuous spectrum of exponents, each one belonging to a given fractal set and leading to a multifractal description of the phenomenon~\cite{benzi1984multifractal}. Similarly, chaotic dynamical systems are often characterized by fractal or multifractal organization in space and in time: given their fully deterministic design, their sensitivity to initial conditions is rather unexpected and counter-intuitive. In this case, one of the emergent features is the lack of predictability above a certain temporal horizon~\cite{Lorenz1963}.

Another broad class of spatio-temporal changes in the concentration of chemical or non-chemical substances can be also captured by reaction-diffusion models, widely used to reproduce the pattern formation -- known also as Turing patterns~\cite{turing1990chemical} -- due to the self-organization of travelling waves. Here, an initially homogeneous substance is locally activated by means of reactions while being inhibited at longer ranges: the competition between these two dynamical processes has been widely used to explain morphogenesis in biology~\cite{meinhardt1982models,turing1990chemical}, chemical reactions~\cite{ouyang1991transition}, epidermal wound healing~\cite{sherratt1990models}, species dynamics~\cite{roques2016modelling} and epidemic spreading within a population~\cite{pastor2015epidemic}. 

Criticality, i.e., the peculiar behavior exhibited by physical systems at critical points which mark the phase transition between qualitatively distinct regimes, provides another reservoir for emergent phenomena: above a critical point the system can exhibit a feature which disappears once the a control parameter, such as the temperature, is tuned below such a threshold. A hallmark of critical phenomena is the presence of long-range correlations within the system' units: they translate into the lack of a characteristic correlation length, which is typical of power laws. Close to the critical point we observe a degree of universality: a small number of scaling exponents that can be used to define universality classes able to describe a broad variety of systems which manifest fractal features~\cite{stanley1972introduction,Wilson1975,wilson1983renormalization,Stanley1999}. A widely known emergent phenomenon, such as ferromagnetism, can be understood in terms of the collective behavior due to the spin-spin interactions of electrons in a material which tend to spontaneously align at the critical temperature, effectively magnetizing the system at large scale. Here, note that ferromagnetism would have no meaning for a system of one particle, since the phenomenon is related to collective behavior causing simultaneous alignment even in absence of an external magnetic field. A paradigmatic approach to gain insight about critical phenomena is the Ising model at different dimensions: it has been shown that a simple two-dimensional Ising model with fields is universal, i.e., it can be used to facilitate the physical simulation of Hamiltonians with complex interactions~\cite{DelasCuevas2016} and it has been related to cellular automata~\cite{domany1984equivalence}. Remarkably, there are systems that do not even need a parameter (e.g., temperature) to be tuned in order to exhibit the scale-invariant organization in space or time, such as in critical systems at phase transition. In fact, such systems dynamically reconfigure their state and spontaneously reach a critical point, which is also an attractor. This peculiar behavior is widely known as self-organized criticality~\cite{Bak1987,bak1988self} (SOC) and it is characteristic of driven nonlinear systems of many interacting units out of equilibrium~\cite{Vespignani1998}: the signature of SOC is the fractal organization in space or time, and it has been observed in biological, ecological, physical and social systems~\cite{Bak1996,Turcotte1999}.

At this point, since most of the systems mentioned so far are non-living, it is worth remarking what it is meant by ``organization'' in open systems out of thermal equilibrium. Here, it is defined by the formation of spatial or temporal (or both) structures that are perceived by an external observer able to measure them in terms of information. On the one hand, this information can be understood as a ``difference which makes a difference''~\cite{Bateson1972}, which does not allow for an operational definition. On the other hand, the mathematical concept of information as introduced by Claude Shannon~\cite{Shannon1948} allows for different observers to define what it is meaningful to them, leading to a subjective definition of which degrees of freedom are relevant for one's description of the system and, consequently, the number of possible states used to calculate Shannon entropy, i.e., the average minimum number of binary digits needed to encode a sequence of symbols. It follows that information depends on the observer~\cite{Adami2016} and, consequently, also the identification of patterns in organizing systems. This is in agreement with the William Ross Ashby's prescription that organization is partly in the eyes of the beholder~\cite{Ashby1947,ashby1962principles}. What can we say, instead, from a thermodynamic perspective? Let us consider a self-organizing system $\mathcal{S}$ and an environment $\mathcal{E}$ defining, once taken together, a closed universe $\mathcal{U}=\mathcal{S}\bigcup\mathcal{E}$. Let $S_\mathcal{S}$ and $S_\mathcal{E}$ denote the entropy of the self-organizing system and the environment, respectively. Of course, $\Delta S_\mathcal{S}/\Delta t > 0$ for a purely thermodynamic non-self-organizing system and $\Delta S_\mathcal{S}/\Delta t = 0$ for a mechanical one. Conversely, self-organization requires that the change of entropy per unit of time is negative: i.e., $\Delta S_\mathcal{S}/\Delta t < 0$ for a sufficient amount of time, which requires the entropy of the environment to change as $\Delta S_\mathcal{E}/\Delta t > 0$ not to violate the second law of thermodynamics for the universe $\mathcal{U}$. The presence of irreversible processes contributing to the decrease of system entropy would be balanced by a larger increase in the rest of the universe, i.e., $\Delta S_\mathcal{U}/\Delta t > 0$. Therefore, from a global perspective, one would be forced to disagree with the definition of self-organization, unless one considers the system to perpetually interact with an environment able to supply energy and order, as pointed out by Heinz von Forster in the '60s~\cite{vanFoerster2003}. A complementary perspective is that dissipative structures in non-living systems, such as flames or hurricanes, are not true organizational systems since inanimate units cannot organize, but just self-order, themselves~\cite{Abel2006}. 

\paragraph{Emergence in living systems.} For living systems, one of the first discussions about self-organization was provided by Erwin Schr\"odinger in the '40s, in an attempt to characterize life from a theoretical physics point of view~\cite{schroedinger1944life}. We can use this perspective as a starting point to briefly review emergent phenomena in living systems, from cells to societies. A heuristic theoretical and computational proof that biological self-organization, or life, is an emergent property of any random dynamical system that possesses a Markov blanket has been given almost a decade ago~\cite{friston2013life}, although a clear mechanism allowing for the transition from an abiotic world to life is still a central problem in research about the origin of life~\cite{Gilbert1986,Maher1988,Martin2008,rasmussen2004transitions,Vasas2010,seoane2018information,Javaux2019}, especially in prebiotic chemistry~\cite{Woos2020,Subramanian2020}, where primordial reaction networks play a fundamental role, as recently pointed out in the case of spontaneous fine-tuning to environment~\cite{horowitz2017spontaneous}. At a higher scale, the spontaneous appearance of multiple cell types and their evolution via ecological context, genomic innovation and/or cooperative integration favored the emergence of multicellular life, boosting biological diversity and complexity: in this context, it has been recently shown that a class of discrete dynamical systems -- known as boolean networks and originally introduced to model gene regulatory systems and reproduce their homeostasis and differentiation~\cite{KAUFFMAN1969} -- can be used to explain cellular differentiation~\cite{marquez2021evolution}. Note that a better understanding of gene regulatory networks, as well as of the protein-protein and the metabolic interactions, might shed light on the mechanistic rules needed to design, synthesize or reconfigure a minimal organism genome~\cite{Hutchison2016}, as well as multicellular organisms and living machines~\cite{Sol2007,Sol2018,Kriegman2020,Levin2020,Blackiston2021}, thus expanding our knowledge of complexity emerging from purely digital systems~\cite{Wilke2001,Lenski2003}.

At a higher scale, the interactions among multicellular organisms lead to unexpected, emergent, phenomena that could not be observed or even defined for a single organism. An emblematic example is the ability of \emph{Physarum polycephalum}, the slime mold, to grow adaptive networks able to solve combinatorial optimization problems even if such an organism lacks a nervous system, the one that it is usually assumed to be necessary for such a purpose~\cite{tero2010rules}. For organisms with a nervous system organized as a neural network, additional properties spontaneously appear -- e.g., from capacity for generalization, to categorization, error correction, and time sequence retention~\cite{hopfield1982neural} -- although some of these features might not be limited to systems of interacting living units.

Other organisms, as social insects, exhibit a level of organization that leads to swarms, a collective behavior that can be explained in terms of active matter far from thermodynamic equilibrium: those collectives enable functions that would be otherwise not accessible by each individual in isolation. Herding worms -- a physically coupled group of individuals which confers mechanofunctional material properties to the collective~\cite{OzkanAydin2021}  -- and ant trails~\cite{chialvo1995swarms} and shimmering honeybees clusters~\cite{peleg2018collective} are emblematic examples of swarms where a large number of interacting individual units effectively behave as a super-organism where information -- from the location of a food source to a migration route -- can be transferred without signalling: remarkably, the larger the group the smaller the needed proportion of individuals driving collective decision-making~\cite{Couzin2005,Sridhar2021} (see~\cite{Couzin2009} for a review). Such an amazing behavior is not limited to insects: from herding~\cite{Hamilton1971} to flocking birds~\cite{Caraco1980,Clark1984} and schooling fish~\cite{Ryer1991,Katz2011,Tunstrm2013}, the formation of ordered structures, multistability, mechanical and energetic efficiency are just a few remarkable features of collective states of large number of individuals which are captured by models grounded on statistical physics~\cite{Bialek2012,vicsek2012collective,Cavagna2013,Mora2016,Cavagna2018}.

\paragraph{Emergence in social systems.} At the scale of humans, interactions among individuals and with the environment are responsible for a variety of emergent phenomena~\cite{ball2004critical}. An emblematic example is provided by social segregation, i.e., the meso-scale organization into clusters, each one characterized by a high level of homophily (e.g., based on gender, ethnicity or socio-economic status). In the '70s, Thomas Shelling proposed a simple mechanistic model to explain the emergence of segregation: a set of individuals, characterized by a feature with at least two distinct flavors, is homogeneously distributed in space. At successive time steps, each individual is allowed to perform a discriminatory choice based on the tolerance to the abundance of individuals with distinct flavor in his/her neighborhood: if this abundance is above a predefined threshold, the individual is left free to randomly move to another location. After some time, clusters of same-flavor individuals spontaneously appear, even in absence of a centralized coordination for their formation~\cite{schelling1971dynamic}.

Another interesting phenomenon is population-scale coordination, or social consensus, where collective behavior spontaneously appear from  the microscopic laws of behavioral contagion despite the absence of a centralized organization~\cite{baronchelli2018emergence}. The spreading of a behavior or of an information shares several features with the spreading of an infectious pathogen: epidemic outbreaks are (often temporary) spontaneous phenomena which exploit social interactions to unfold, clustering in space and time~\cite{pastor2015epidemic}. Similarly, a traffic jam cannot be defined at the level of a single unit: be pedestrians or vehicles, whose dynamic is constrained (or not) to follow lanes and directions, different kinds of congestion usually occur well before the road capacity is reached and such a behavior can be partially reproduced by microscopic (particle-based), mesoscopic (gas-kinetic), and macroscopic (fluid-dynamic) models~\cite{helbing2001traffic}. 

Finally, it is worth mentioning another class of large-scale fascinating complex systems: cities. They are complex from many point of views, consisting of a many sub-systems, such as social, economic, environmental ones and their combination. It has been shown that a small set of basic principles, operating at a local level, are enough to explain the growth, the large-scale regularities and the scaling laws observed in cities~\cite{bettencourt2007growth,bettencourt2013origins,barthelemy2016structure}, once again well captured by models grounded on statistical physics~\cite{barthelemy2019statistical,verbavatz2020growth}.

\paragraph{The role of complex networks in emergent phenomena.} A large class of complex systems is characterized by a structure that can be represented in terms of units interconnected by links, which encode one or more kind of interactions or relationships~\cite{boccaletti2006complex}. Mathematically, such systems can be represented by a matrix or, for more complex systems characterized by multiple types of relationships simultaneously -- such as multilayer networks -- by a tensor~\cite{de2013mathematical,artime2022multilayer}. 

On the one hand, the network backbone itself can represent an emergent feature $\Sigma$ under some conditions or constraints. This is the case of nested interaction networks that are the result of an optimization principle which maximizes the abundance of species in mutualistic communities~\cite{suweis2013emergence}.

On the other hand, the network backbone can be considered as the starting point for the analysis of structural and dynamical properties $\Pi$ of a complex system. A pioneering work in this direction, unraveled the emergence of power-law scaling in the connectivity distribution of a variety of networks, from biological to technological ones~\cite{barabasi1999emergence}. This discovery led to a plethora of fundamental insights about the behavior of interconnected systems, from their extreme fragility to targeted attacks to their role in explosive phenomena~\cite{d2019explosive} (see further in this section). 

One of the most striking -- and ubiquitous -- features of empirical complex networks is the emergence of a mesoscale organization, such as hierarchical~\cite{Simon1962,Simon1977} and/or modular~\cite{newman2012communities,fortunato2016community,Peixoto2019} structure,  which has been linked to efficiency in information exchange, functional segregation and integration~\cite{ravasz2002hierarchical,ravasz2003hierarchical,han2004evidence,guimera2005functional,colizza2006detecting,hintze2008evolution,taylor2009dynamic,bullmore2012economy,ghavasieh2020statistical,bertagnolli2021quantifying}. Networks exhibit other emergent features, such as latent geometry~\cite{boguna2021network} or distinct flavors of multilayer organization, from interdependence~\cite{gao2012networks} to multiplexity~\cite{mucha2010community,de2013mathematical,boccaletti2014structure,kivela2014multilayer,de2016physics,artime2022multilayer}. 

Uncovering structural features of networks is a necessary step towards understanding the function(s) of the underlying complex system they are representing, since functionality interlaces with the dynamics of or on the network. For instance, the network counterpart of the Anderson localization has been reported~\cite{burda2009localization}. In the case of networks of oscillators, the collective phenomenon of synchronization spontaneously emerges if the coupling between units is above a critical threshold~\cite{kuramoto1975international,acebron2005kuramoto,arenas2008synchronization,rodrigues2016kuramoto}, with a variety of phenomena ranging from explosive behavior in scale-free networks~\cite{gomez2011explosive} to new types of collective states emerging from coupling synchronization dynamics with swarming behavior, like in swarmalators~\cite{o2017oscillators}, or network's dynamics~\cite{ghosh2022synchronized}. Similarly, some critical properties -- such as the emergence of metacritical points -- start to depend on the way distinct dynamics are coupled together, such as in interacting spreading phenomena on the top of simple or multilayer networks~\cite{granell2013dynamical,sanz2014dynamics,de2016physics,castioni2021critical}.

As mentioned above, the network representation of empirical systems usually supports dynamical processes on them~\cite{barrat2008dynamical}. For instance, in the power grid the electricity is generated and delivered, transportation networks sustain a flow of people and goods from one place to another, users navigate the content in the World Wide Web through hyperlinks, information or a pathogen spreads in online and offline social networks via friendship and acquaintance ties, etc. To some extent, the sustained network-wide functionality can be seen as a robust emergent phenomenon that dodges disrupting events -- such as errors, random failures or attacks -- in the individual units of the network. Thus, functionality, robustness and resilience can be seen as complementary emerging properties of a system.

The link between network functionality and robustness has been actively investigated due to its societal impact, for instance, at the infrastructural or ecological levels. Bare-bones approaches have looked at the size of the largest connected component of the network, assuming it is the most functional part, when the original system is perturbed with the removal of a given fraction of nodes or links. This is the reversed process of adding nodes or links in an initially empty network and track when a macroscopically functional structure emerges. Both processes are completely equivalent in the absence of hysteresis loops. Percolation theory~\cite{li2021percolation} turns out to be useful in this case, as it provides a set of concepts and analytical and computational techniques suitable to describe this functional-to-nonfunctional transition. Diverse intervention protocols have been proposed to dismantle the system: random uniform selection of nodes to model unexpected disruptions~\cite{albert2000error}, while informed interventions can be seen as targeted attacks. Examples of the latter included making use of both topological~\cite{albert2000error,holme2002attack,grassia2021machine} and non-topological information~\cite{artime2021percolation}. Moreover, the emergence of the functional structure can be achieved in an explosive, abrupt manner via the design of topologically sophisticated rules~\cite{goltsev2006k,baxter2010bootstrap,achlioptas2009explosive} or via the inclusion of interdependencies~\cite{son2012percolation,radicchi2015percolation}.

Another realistic approach is to consider how networks respond to cascading failures. These processes are characterized by an initial stressor -- located in a small region of the network -- that is able to spread and impact large portions of the network. Different propagation rules have been proposed, whose characteristics depend on the system one is trying to model (see~\cite{watts2002simple,motter2002cascade,brummitt2012suppressing,yu2016system,huang2013cascading,shekhtman2016recent}). Of particular interest are the cascades spreading in multilayer and interdependent structures, as it has been shown that they could suddenly collapse, thus making difficult to identify early signals of fragmentation~\cite{buldyrev2010catastrophic,gao2012networks}. Instead of a malfunction that spreads, one can also consider how the exploration or navigability properties of walkers are affected when some of the network components are corrupted. It has been reported that these characteristics are greatly impacted by the walk strategy and topological properties~\cite{de2014navigability}.

%%%%%%%%%%%%%%%%%%%%%%%%%%%%%%%%%

\section{Defining emergence from a mathematical perspective}

David Chalmers clearly distinguishes between two types of emergence: \emph{weak} and \emph{strong}. A phenomenon at a high scale is \emph{weakly emergent} with respect to a lower scale if, given the laws governing the latter, the patterns observed in the former are unexpected, but they can be deducible in principle from advanced calculations and/or computation. If such a deduction is not possible even in principle, then the phenomenon is \emph{strongly emergent}~\cite{chalmers2006}. A similar distinction was already present in the work of Mark Bedau, who identified two hallmarks of emergence where phenomena are either i) somehow constituted by, and generated from, underlying processes; or ii) somehow autonomous from underlying processes~\cite{bedau1997weak}. The existence of strongly emergent phenomena would require new fundamental laws of nature for their explanation. In fact, it is argued that only the weak emergence is scientific relevant, consistent with materialism and metaphisically innocent to provide a ground for a science of complexity~\cite{bedau1997weak}. The possibility that a simple initial configuration might evolve into unexpected patterns allows to overcome the reductionist approaches while preserving the possibility for rich phenomena at distinct levels of explanation and, consequently, an ultimately physicalist picture of the world~\cite{chalmers2006}. We capitalize on these arguments to rationalize an operational definition.

\begin{figure*}[!ht]
\centering
\includegraphics[width=\textwidth]{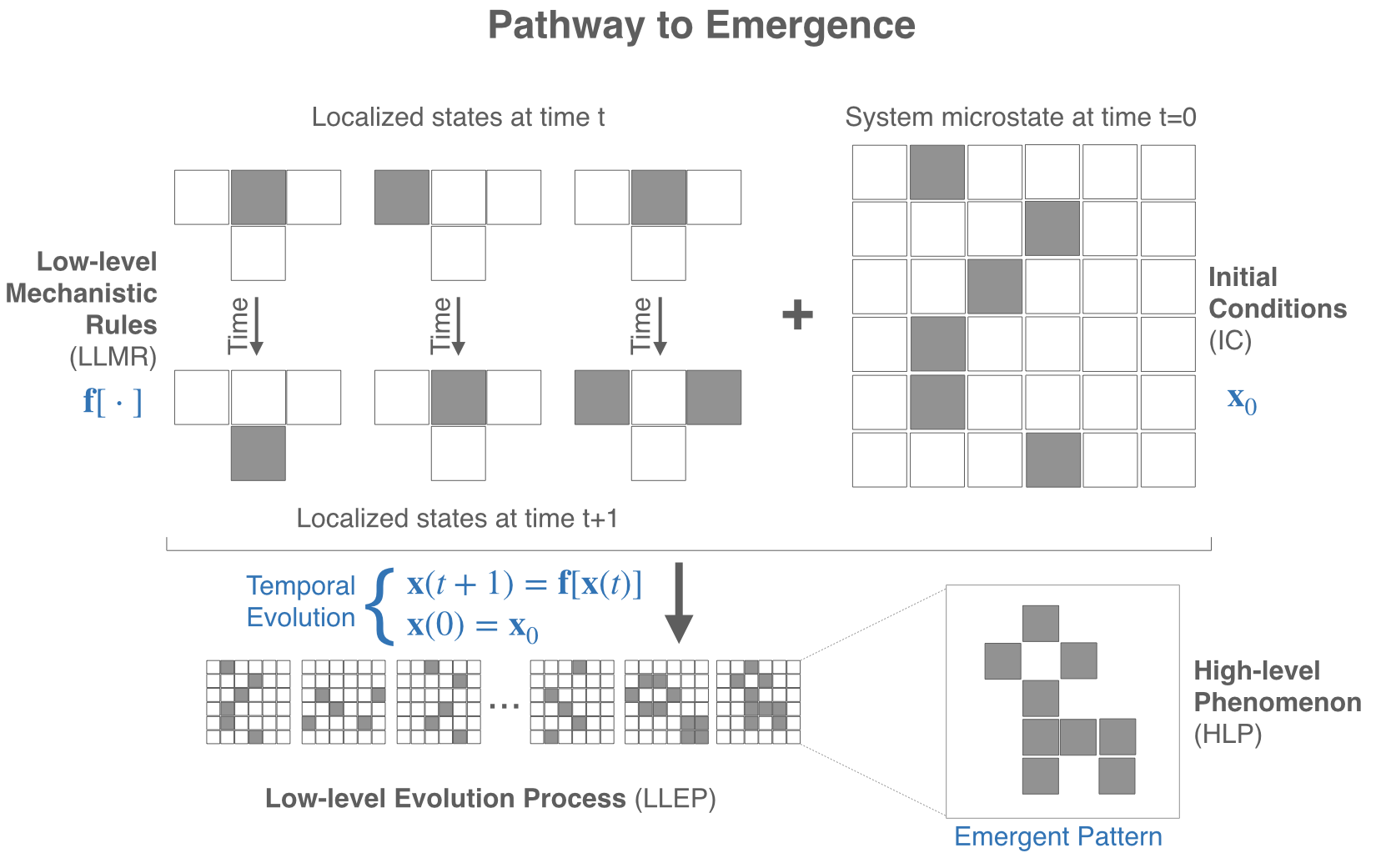}
\caption{\label{fig:fig1} A complex system consists of interconnected units for which a set of local mechanistic rules for hanging over time is assigned (top-left). An initial condition for such a system is given (top-right) and the system evolves according to its microscopic rules until an emergent pattern is observed. The reader might identify an analogy with the well-known Game of Life, a cellular automaton proposed by John Conway in the '70s~\cite{gardner1970fantastic} for which, more recently, quantum versions are being explored~\cite{ney2022entanglement}. Note, however, that we do not require the state of the units to be binary, or even discrete. The analogy can be used to better understand the rich basin of phenomena that can take place when a few microscopic rules and adequate initial conditions are considered: in fact, it should be noted that not all possible LLMR and IC lead to HLP. See the text for further details.}
\end{figure*}

Let a system $\mathcal{S}$ be made up of a finite number of many units $u$ that interact among them and/or with the environment. Emergence is the apparition of system-wide properties or qualities that are not present individually in the units but have their origin precisely in the interactions. Let us assign a set of simple low-level mechanistic rules (LLMR) $\mathbf{f[\cdot]}$ that allow our units to change locally. If we indicate the state of the system at a specific time $t$ by $\mathbf{x}(t)$, then such rules can be encoded into an evolution equation that can be discrete or continuous in time. For simplicity, let us consider a discrete-time evolution, as schematically shown in Fig.~\ref{fig:fig1}, and an initial system state which is mathematically represented by
\begin{equation}
    \begin{cases}
      \mathbf{x}(t+1)= & \mathbf{f}[\mathbf{x}(t)]\\
      \mathbf{x}(t=0)= & \mathbf{x}_{0}
    \end{cases}       
\end{equation}
where $\mathbf{x}_{0}$ defines initial conditions (IC). The microscopic evolution of the system is defined as low-level evolution process (LLEP) and, after some time, it will lead to a stable or metastable emergent pattern, i.e., a high-level phenomenon (HLP). Note that we did not specify if $\mathbf{f}[\cdot]$ is deterministic or stochastic, and if the system  is open or closed, since any of their combination is plausible, in principle. In fact, if the system is open and the dynamical rules are deterministic, the contingencies of the flux of parts and states through $\mathcal{S}$ provide additional external conditions, whereas if the system is closed there is only one external condition, i.e.,  the initial one. If the dynamics is stochastic, then accidental effects provide additional external conditions. It is clear as a full knowledge of $\mathcal{S}$, $\mathbf{f}[\cdot]$, $\mathbf{x}_{0}$ and eventual external conditions might allow, through computational analysis, to evolve microstates and observe unexpected macrostates. Summarizing:

\begin{itemize}
\item\textbf{Non-emergent phenomena:} knowledge of LLMR and IC allows to deduce expected HLP;
\item\textbf{Weakly emergent phenomena:} knowledge of LLMR and IC allows to deduce unexpected HLP through computation (e.g., simulations);
\item\textbf{Strongly emergent phenomena:} knowledge of LLMR and IC does not allow to deduce HLP even in principle.
\end{itemize}

Note that the above classification applies well for two distinct types of emergent phenomena we have been discussing: 1) the ones where the system $\Sigma$ itself emerges from its constituents, as for the spontaneous organization into structural backbones mentioned in the previous section; 2) the ones where macroscopic properties $\Pi$ emerge because of the existence of a system (e.g., superconductivity or the robustness of a network to random disruptions). Such a hierarchy between emergent phenomena, involving large-scale structures and properties, can be used to better understand the results of the studies published in this Theme Issue.

\section{Summary of the Theme Issue}

At this point, the general reader should be familiar with the concept of emergence. In this section, we briefly introduce the papers in this collection: each manuscript has been developed independently from the others but at the same time it is connected to them to allow for the exploration of emergence from a multidisciplinary and interdisciplinary perspective. To some extent, the emergent result is this Theme Issue. They are organized, here and in the Issue, as follows: we first present the theoretical contributions dealing with the problem of emergence itself, then we move to the contributions about the emergence of specific phenomena in different contexts. The latter are introduced according to their scale, from smallest (quantum realm) to largest ones (epidemics). 

The Theme Issue starts with the article \textbf{Emergence and algorithmic information dynamics of systems and observers} by Abrah\~{a}o and Zenil~\cite{abrahao2022emergence}. The authors deal with the problem of asserting whether or not a phenomenon can be considered emergent, from the viewpoint of computation theory. Identifying the act of observing as a mutual perturbation between the system and the observer, they find that the emergence of algorithmic information is dependent on the observer's formal knowledge but robust to other subjective factors. Additionally, they prove that emergence becomes observer-independent if it increases in an unbounded and rapid fashion, and two examples are studied to illustrate this phenomenon. 

It follows the article by Rosas and collaborators, \textbf{Greater than the parts: A review of the information decomposition approach to causal emergence}~\cite{rosas2022greater}, where it is offered an accessible and rigorous review of a recently developed formal theory of causal emergence. This theory is based on information decomposition, where emergence is considered a property of part-whole relationships within the system under study. They present a mathematical background, the key principles of the theory and several case studies, both from empirical data and synthetic simulations, that demonstrate the applicability of their approach.

In the search for a solid and convincing theory of emergence, Varley and Hoel's contribution, \textbf{Emergence as the conversion of information: A unifying theory}~\cite{varley2022emergence}, overcomes the traditional dichotomy between strong and weak emergence and, aiming at finding a formal theory of emergence able to identify the intrinsic scale of function of complex systems, propose a mathematical framework in which emergence is identified with information conversion across scales. They base it on information theory and successfully apply it on a model system of Boolean networks.

In \textbf{Emergence of functional information from multivariate correlations}~\cite{adami2022emergence}, Adami and Nitash draw a mapping between the multivariate correlations within a symbolic sequence, such as the nucleic- or aminoacid ones, and the functional information encoded on it. Their model-free approach is tested in the largest known computational genotype-phenotype map, in which they are successful in distinguishing functional from non-functional sequences.

We have already discussed that one necessary condition to find emergence is to have a system made of many small interacting sub-systems. In \textbf{Strengthened second law for multi-dimensional systems coupled to multiple thermodynamic reservoirs}~\cite{wolpert2022strengthened}, David Wolpert, studies the stochastic thermodynamic properties of such systems, under the only requirement that they evolve according to a continuous-time Markov chain. Lower bounds for the entropy production are derived, hence offering a strengthened version of the second law of thermodynamics. 

The critical phenomena observed in phase transitions are a paradigmatic example of how local interactions lead to system-wide effects. In their article \textbf{Emergent entanglement structures and self-similarity in quantum spin chains}~\cite{sokolov2022emergent}, Sokolov \textit{et al}. provide a thorough characterization of this in the quantum realm. By means of tools borrowed from complex network theory, they unveil new emergent phenomena in spins interacting through the XX model of magnetism, such as an entangled community structure, topological instabilities and self-similarity in the entanglement network.

Moving up to higher scales, we pass from spins to molecules and metabolism. In Nader, Sebastianelli and Mansy’s Opinion Piece, \textbf{Protometabolism as out-of-equilibrium chemistry}~\cite{nader2022protometabolism}, the authors argue for the need of exploring the role played by prebiotic energy sources as a possible explanation of the origin of metabolism. They put the focus on the out-of-equilibrium chemical properties, advocating for non-hydrothermal vents as regions of the early Earth that could be able to provide the necessary energy to sustain the first protocells.

Still at the molecular scale, Xavier and Kauffman's article \textbf{Small-molecule autocatalytic networks are universal metabolic fossils}~\cite{xavier2022small}  focuses on the emergence of early metabolism. They investigate small-molecule reflexively autocatalytic food-generated networks, proving that these structures can be generated from all the hitherto annotated prokaryotic metabolic networks in the KEGG database. The results based on the analysis of these networks yields the striking conclusion that molecular reproduction started much earlier than the last universal common ancestor. 

Molecular function emerges from molecular evolution. In the review article \textbf{The simple emergence of complex molecular function}~\cite{manrubia2022simple}, Manrubia guides us through some of the mechanisms that facilitate this evolution, such as phenotypic bias, genotype-to-phenotype redundancy, among others. When taken together all these mechanisms, molecular complexity seems the most natural outcome.

Moving to larger scales, we arrive to neurons and the brain. It is well-known that brain states, both healthy and altered, can be characterized by complex emergent spatio-temporal patterns. In \textbf{Understanding brain states across spacetime informed by whole-brain modelling}~\cite{vohryzek2022understanding} Vohryzek and coauthors embrace the idea of considering the human brain as a complex system, and they offer a review on how these patterns can be mapped and modeled via non-invasive imaging and whole-brain modeling, with a focus on depression and psychedelics.

In the brain, we also find synchronisation phenomena. Buend\'{i}a and coauthors provide in \textbf{The broad edge of synchronisation: Griffiths effects and collective phenomena in brain networks}~\cite{buendia2022broad} a thorough characterization of the rich dynamical repertoire that arises in brain synchronization when it is combined a minimal dynamical model of neural activity with empirically-observed properties of brain connectivity, such as hierarchical-modular and core-periphery structures. They reveal the emergence of complex collective states with flexible levels of synchronization, which is a necessary step towards a better understanding of the functional capabilities of brains.

Saeedian \textit{et al}.'s article \textbf{Effect of delay on the emergent stability patterns in Lotka-Volterra ecological dynamics}~\cite{saeedian2022effect} finds itself at the ecosystem scale. They tackle the problem of ecosystem stability when the realistic feature of delayed interactions between species is incorporated in a generalized Lotka-Volterra model. They provide analytical and numerical results as a function of the delay strength, and report a detrimental effect on the ecosystem stability as delay increases. At a critical value of the delay, oscillatory states emerge, which is a dynamical regime that could not be predicted by standard linear stability analysis.
    
Peters and Adamou tackle the problem of cooperation, understood as resource sharing between system units, such as cells, animals, humans, institutions, etc. In the Opinion Piece \textbf{The ergodicity solution of the cooperation puzzle}~\cite{peters2022ergodicity}, they propose a simple model where this behavior, which might seem not appealing to one of the cooperators, can arise even if the classical assumptions of reciprocity or the existence of net benefit between cooperators are not met, as far as the resources follow a noisy multiplicative growth. Thus their model becomes a candidate to explain cooperation in many real setting and provides a baseline for behavioral comparison.

Our journey across scale reaches the human one at this point. In the review \textbf{A research agenda for the study of social norm change}~\cite{andrighetto2022research},  
Andrighetto and Vriens provide a complete overview on social norm change. Indeed, interactions among social agents are the basis on which norms emerge and develop, hence being an interesting tool to tackle collective action problems. They critically discuss how to identify social norms, how to establish causal effects, how norm change is linked to tipping point dynamics, and outline future research problems.

Humans do not only interact directly among us, but also with technological devices. Brinkmann and coauthors shed light on the role that the interaction between humans and algorithms might play in shaping the emergent properties of cultural evolution. In \textbf{Hybrid social learning in human-algorithm cultural transmission}~\cite{brinkmann2022hybrid} they propose a set of 6 hypothesis, related to the improvement of collective performance tasks via (hybrid) social learning, that are tested in an experimental set-up. Their empirical findings highlight the importance of biases: even if an algorithm aims to aid humans, the provided  information can be quickly lost in successive human-human interactions due to, precisely, human biases.

Closing the Theme Issue, we find the article \textbf{Emergence of protective behaviour under different risk perceptions to disease spreading}~\cite{khanjanianpak2022emergence}, by Khanjanianpak \textit{et al}. The authors address the problem of how different behavioral responses emerge when a population is exposed to a given risk, when the latter is heterogeneously perceived within the population. They focus on the timely problem of the adoption of protective measures, such as social distancing, social distancing, etc., during the course of a disease spreading.

\section{Conclusions and Outlooks}

Since the first discovery of emergent phenomena a plethora of phenomenological evidence has been provided to show that they are ubiquitous, from quantum to classical physical systems, from non-living to living ones. Nevertheless, we envision outstanding challenges and a rather exciting agenda for both fundamental and applied research on emergence in the next future. 

On the one hand, many emergent processes exhibit similar properties -- e.g., cluster formation in critical physical systems consisting of many particles, and in social processes -- and domain-specific peculiarities. Unraveling the building blocks of an emergent phenomenon in space, time, or both in space and time, is still an open problem: while similarities across disciplines suggest the existence of a few generative rules, thus increasing the likelihood of finding such mechanisms, the advances in this direction might be slowed down by domain-specific microscopic rules which might be difficult to reconcile within a comprehensive vision. It is worth remarking here that it is still not granted that such a comprehensive picture exists or if it useful at all. Moreover, even the inverse problem is an open challenge: predicting the macroscopic outcome(s) of a process, given the microscopic rules governing system's units might lead to the discovery of a new kind of universality classes. Recent advances in the latter are related to programmable pattern formation, for which there are interesting applications to the case of cellular systems with local signaling~\cite{Ramalho2021}.

On the other hand, the outcome of such an understanding at a fundamental level might open the door to countless applications in physics, biology and engineering, to mention a few ones. One might be able to design physical systems exhibiting (weakly) emergent properties, such as robot swarms able to self-assembly~\cite{Rubenstein2014}, self-repair and exhibit high robustness to internal failures~\cite{Li2019}, with the ultimate goal to engineer complex systems able to perform specific tasks and achieve human-designed goals. The basin of applications ranges from medicine to cybersecurity.

In a nutshell, in such robot swarms each unit communicates or interact only locally with its neighbors: all together they are able to generate flexible and scalable collective behaviors without relying on external infrastructures or centralized control, actively adapting in response to stimuli from the environment in which they are embedded. The interested reader is referred to~\cite{brambilla2013swarm}. 

An emblematic example is given by large flocks of autonomous drones that seamlessly navigate in confined spaces~\cite{Vsrhelyi2018}. Furthermore, it has been shown that robophysical systems known as ``smarticles'' -- planar ensembles of periodically deforming smart, active particles -- are able to generate endogenous phototaxis, a kind of locomotory movement occurring when a collective of organisms moves in response to presence or absence of light, thus providing a model to develop internal mechanical interactions to perform tasks without a centralized coordination~\cite{Savoie2019}. As biological systems combine microscopic stochastic components to achieve a desired macroscopic function, such as cell migration in morphogenesis, tissue repair, and cancer~\cite{haeger2015collective}, robot swarms might achieve a similar behavior. Recently, it has been shown that robot swarms are able to mimic the behavior of labour division in ant colonies~\cite{Aguilar2018} and herding worms~\cite{OzkanAydin2021}, undergoing shape transformations which make the system more robust to thermal stress or more energetically efficient. 

In computer science, swarm learning, a machine learning explicitly based on decentralized approaches that rely on networks of learners, has outperformed the standard federating learning. Remarkably, it has been shown that it achieves better results than scenarios in which each node in the network learns separately~\cite{WarnatHerresthal2021}.

Such exciting advances in technological applications of self-organizing artificial systems, at both software and hardware level, might be employed to detect threats in IT systems and build robust security layers, as well as to accelerate the recovery of networked systems and infrastructures -- e.g., telecommunications, power, water management, supply chain, so forth and so on -- or reduce to gap to precision medicine with personalized clinical treatments.

Wrapping up, there is a great promise in unraveling the principles of emergent phenomena which could be potentially find groundbreaking application in material science, nanotechnology, medicine, engineering and computer science. If we are forced to summarize the concept of emergence by means of figurative language, we can safely assess that there were no lasagne encoded in the Big Bang.

\vskip6pt
\enlargethispage{20pt}

\ethics{NA}

\dataccess{NA}

\aucontribute{The authors contributed equally to this work.}

\competing{The authors declare no competing interests.}

\funding{NA}

\ack{The authors thank Ricard Sol\'e for feedbacks about this manuscript, as well as Andrea Cavagna, Gemma De las Cuevas, Artemy Kolchinsky, Angelo Vulpiani, Sarah Imari Walker and Joana Xavier, for stimulating discussions on emergence, during the Complexity Webinars series organized by the CoMuNe Lab in 2021. We are also doubtlessly indebted to all the authors whose contributions give shape to the present Issue, as well as to all the referees that have contributed with their timely reviews and constructive comments. We also gratefully acknowledge the Commission Editor, Alice Power, for her constant support and help throughout.} 

\disclaimer{NA}

%%%%%%%%%% Insert bibliography here %%%%%%%%%%%%%%
\bibliographystyle{RS}
\bibliography{biblio.bib}

\begin{thebibliography}{99}

\bibitem{cox2005complex}
Cox DL, Pines D. 2005  Complex adaptive matter: emergent phenomena in
  materials. {\em MRS bulletin} \textbf{30}, 425--432.

\bibitem{koffka1935gestalt}
Koffka K. 1935 {\em Principles of gestalt psychology}.
New York: Harcourt, Brace and Company.

\bibitem{wheeler1926emergent}
Wheeler WM. 1926  Emergent evolution and the social. {\em Science} \textbf{64},
  433--440.

\bibitem{wheeler1927emergent}
Wheeler W. 1927 {\em Emergent Evolution and the Social}.
Number No. 11 in Emergent Evolution and the Social. K. Paul, Trench, Trubner \&
  Company, Limited.

\bibitem{von1950outline}
von Bertalanffy L. 1950  An Outline of General System Theory. {\em The British
  Journal for the Philosophy of Science} \textbf{1}, 134--165.

\bibitem{ludwig1951general}
Ludwig~von Bertalenffy K. 1951  General system theory--A new approach to unity
  of science. {\em Human Biology} \textbf{23}, 303--361.

\bibitem{anderson1972more}
Anderson PW. 1972  More is different. {\em Science} \textbf{177}, 393--396.

\bibitem{nicolis1977self}
Nicolis G, Prigogine I. 1977 {\em Self-organization in nonequilibrium systems:
  From Dissipative Structures to Order through Fluctuations}.
Wiley.

\bibitem{bedau2008emergence}
Bedau MA, Humphreys PE. 2008 {\em Emergence: Contemporary readings in
  philosophy and science.}
MIT press.

\bibitem{wolfram1983statistical}
Wolfram S. 1983  Statistical mechanics of cellular automata. {\em Reviews of
  Modern Physics} \textbf{55}, 601.

\bibitem{cook2004universality}
Cook M et~al.. 2004  Universality in elementary cellular automata. {\em Complex
  Systems} \textbf{15}, 1--40.

\bibitem{laughlin1999nobel}
Laughlin RB. 1999  Nobel lecture: Fractional quantization. {\em Reviews of
  Modern Physics} \textbf{71}, 863.

\bibitem{gellmann1994the}
Gell-Mann M. 1994 {\em The quark and the jaguar: Adventures in the simple and
  the complex}.
New York: Owl Books.

\bibitem{de2019complexity}
De~Domenico M, Brockmann D, Camargo C, Gershenson C, Goldsmith D, Jeschonnek S,
  Kay L, Nichele S, Nicol{\'a}s J, Schmickl T et~al.. 2019  Complexity
  explained. .

\bibitem{wilczek2012origins}
Wilczek F. 2012  Origins of mass. {\em Central European Journal of Physics}
  \textbf{10}, 1021--1037.

\bibitem{anderson1958absence}
Anderson PW. 1958  Absence of diffusion in certain random lattices. {\em
  Physical Review} \textbf{109}, 1492.

\bibitem{bardeen1957microscopic}
Bardeen J, Cooper LN, Schrieffer JR. 1957  Microscopic theory of
  superconductivity. {\em Physical Review} \textbf{106}, 162.

\bibitem{SpecialIssue2020}
Editorial. 2020  Emergent superconductivity. {\em Nature Physics} \textbf{16},
  705--705.

\bibitem{josephson1974discovery}
Josephson BD. 1974  The discovery of tunnelling supercurrents. {\em Reviews of
  Modern Physics} \textbf{46}, 251.

\bibitem{Klitzing1980}
v.~Klitzing K, Dorda G, Pepper M. 1980  New Method for High-Accuracy
  Determination of the Fine-Structure Constant Based on Quantized Hall
  Resistance. {\em Physical Review Letters} \textbf{45}, 494--497.

\bibitem{yennie1987integral}
Yennie D. 1987  Integral quantum Hall effect for nonspecialists. {\em Reviews
  of Modern Physics} \textbf{59}, 781.

\bibitem{laughlin1981quantized}
Laughlin RB. 1981  Quantized Hall conductivity in two dimensions. {\em Physical
  Review B} \textbf{23}, 5632.

\bibitem{tsui1982two}
Tsui DC, Stormer HL, Gossard AC. 1982  Two-dimensional magnetotransport in the
  extreme quantum limit. {\em Physical Review Letters} \textbf{48}, 1559.

\bibitem{laughlin1983anomalous}
Laughlin RB. 1983  Anomalous quantum Hall effect: an incompressible quantum
  fluid with fractionally charged excitations. {\em Physical Review Letters}
  \textbf{50}, 1395.

\bibitem{ahlers2009heat}
Ahlers G, Grossmann S, Lohse D. 2009  Heat transfer and large scale dynamics in
  turbulent Rayleigh-B{\'e}nard convection. {\em Reviews of Modern Physics}
  \textbf{81}, 503.

\bibitem{swinney1978hydrodynamic}
Swinney HL. 1978  Hydrodynamic instabilities and the transition to turbulence.
  {\em Progress of Theoretical Physics Supplement} \textbf{64}, 164--175.

\bibitem{benzi1984multifractal}
Benzi R, Paladin G, Parisi G, Vulpiani A. 1984  On the multifractal nature of
  fully developed turbulence and chaotic systems. {\em Journal of Physics A:
  Mathematical and General} \textbf{17}, 3521.

\bibitem{Lorenz1963}
Lorenz EN. 1963  Deterministic Nonperiodic Flow. {\em Journal of the
  Atmospheric Sciences} \textbf{20}, 130--141.

\bibitem{turing1990chemical}
Turing AM. 1990  The chemical basis of morphogenesis. {\em Bulletin of
  Mathematical Biology} \textbf{52}, 153--197.

\bibitem{meinhardt1982models}
Meinhardt H. 1982  Models of biological pattern formation. {\em New York}
  \textbf{118}.

\bibitem{ouyang1991transition}
Ouyang Q, Swinney HL. 1991  Transition from a uniform state to hexagonal and
  striped Turing patterns. {\em Nature} \textbf{352}, 610--612.

\bibitem{sherratt1990models}
Sherratt JA, Murray JD. 1990  Models of epidermal wound healing. {\em
  Proceedings of the Royal Society of London. Series B: Biological Sciences}
  \textbf{241}, 29--36.

\bibitem{roques2016modelling}
Roques L, Bonnefon O. 2016  Modelling population dynamics in realistic
  landscapes with linear elements: A mechanistic-statistical reaction-diffusion
  approach. {\em PloS one} \textbf{11}, e0151217.

\bibitem{pastor2015epidemic}
Pastor-Satorras R, Castellano C, Van~Mieghem P, Vespignani A. 2015  Epidemic
  processes in complex networks. {\em Reviews of Modern Physics} \textbf{87},
  925.

\bibitem{stanley1972introduction}
Stanley HE, Wong VK. 1972  Introduction to phase transitions and critical
  phenomena. {\em American Journal of Physics} \textbf{40}, 927--928.

\bibitem{Wilson1975}
Wilson KG. 1975  The renormalization group: Critical phenomena and the Kondo
  problem. {\em Reviews of Modern Physics} \textbf{47}, 773--840.

\bibitem{wilson1983renormalization}
Wilson KG. 1983  The renormalization group and critical phenomena. {\em Reviews
  of Modern Physics} \textbf{55}, 583.

\bibitem{Stanley1999}
Stanley HE. 1999  Scaling, universality, and renormalization: Three pillars of
  modern critical phenomena. {\em Reviews of Modern Physics} \textbf{71},
  S358--S366.

\bibitem{DelasCuevas2016}
De~las Cuevas G, Cubitt TS. 2016  Simple universal models capture all classical
  spin physics. {\em Science} \textbf{351}, 1180--1183.

\bibitem{domany1984equivalence}
Domany E, Kinzel W. 1984  Equivalence of cellular automata to Ising models and
  directed percolation. {\em Physical review letters} \textbf{53}, 311.

\bibitem{Bak1987}
Bak P, Tang C, Wiesenfeld K. 1987  Self-organized criticality: An explanation
  of the 1/fnoise. {\em Physical Review Letters} \textbf{59}, 381--384.

\bibitem{bak1988self}
Bak P, Tang C, Wiesenfeld K. 1988  Self-organized criticality. {\em Physical
  Review A} \textbf{38}, 364.

\bibitem{Vespignani1998}
Vespignani A, Dickman R, Mu{\~{n}}oz MA, Zapperi S. 1998  Driving,
  Conservation, and Absorbing States in Sandpiles. {\em Physical Review
  Letters} \textbf{81}, 5676--5679.

\bibitem{Bak1996}
Bak P. 1996 {\em How Nature Works}.
Springer New York.

\bibitem{Turcotte1999}
Turcotte DL. 1999  Self-organized criticality. {\em Reports on Progress in
  Physics} \textbf{62}, 1377--1429.

\bibitem{Bateson1972}
Bateson G. 1972 {\em Steps to an Ecology of Mind}.
University of Chicago Press.

\bibitem{Shannon1948}
Shannon CE. 1948  A Mathematical Theory of Communication. {\em Bell System
  Technical Journal} \textbf{27}, 379--423.

\bibitem{Adami2016}
Adami C. 2016  What is information?. {\em Philosophical Transactions of the
  Royal Society A: Mathematical, Physical and Engineering Sciences}
  \textbf{374}, 20150230.

\bibitem{Ashby1947}
Ashby WR. 1947  Principles of the Self-Organizing Dynamic System. {\em The
  Journal of General Psychology} \textbf{37}, 125--128.

\bibitem{ashby1962principles}
Ashby WR. 1962  Principles of the self-organizing system. In v.~Foerster H,
  Zopf GW, editors, {\em Principles of Self-Organization: Transactions of the
  University of Illinois Symposium} pp. 255--278. London: Pergamon.

\bibitem{vanFoerster2003}
Von~Foerster H. 2003  On self-organizing systems and their environments. In
  {\em Understanding Understanding} pp. 1--19. Springer, New York, NY.

\bibitem{Abel2006}
Abel DL, Trevors JT. 2006  Self-organization vs. self-ordering events in
  life-origin models. {\em Physics of Life Reviews} \textbf{3}, 211--228.

\bibitem{schroedinger1944life}
Schr\"odinger E. 1944 {\em What is life}.
Cambridge University Press.

\bibitem{friston2013life}
Friston K. 2013  Life as we know it. {\em Journal of the Royal Society
  Interface} \textbf{10}, 20130475.

\bibitem{Gilbert1986}
Gilbert W. 1986  Origin of life: The {RNA} world. {\em Nature} \textbf{319},
  618--618.

\bibitem{Maher1988}
Maher KA, Stevenson DJ. 1988  Impact frustration of the origin of life. {\em
  Nature} \textbf{331}, 612--614.

\bibitem{Martin2008}
Martin W, Baross J, Kelley D, Russell MJ. 2008  Hydrothermal vents and the
  origin of life. {\em Nature Reviews Microbiology} \textbf{6}, 805--814.

\bibitem{rasmussen2004transitions}
Rasmussen S, Chen L, Deamer D, Krakauer DC, Packard NH, Stadler PF, Bedau MA.
  2004  Transitions from nonliving to living matter. {\em Science}
  \textbf{303}, 963--965.

\bibitem{Vasas2010}
Vasas V, Szathmary E, Santos M. 2010  Lack of evolvability in self-sustaining
  autocatalytic networks constraints metabolism-first scenarios for the origin
  of life. {\em Proceedings of the National Academy of Sciences} \textbf{107},
  1470--1475.

\bibitem{seoane2018information}
Seoane LF, Sol{\'e} RV. 2018  Information theory, predictability and the
  emergence of complex life. {\em Royal Society Open Science} \textbf{5},
  172221.

\bibitem{Javaux2019}
Javaux EJ. 2019  Challenges in evidencing the earliest traces of life. {\em
  Nature} \textbf{572}, 451--460.

\bibitem{Woos2020}
Wo{\l}os A, Roszak R, {\.{Z}}{\k{a}}d{\l}o-Dobrowolska A, Beker W,
  Mikulak-Klucznik B, Sp{\'{o}}lnik G, Dygas M, Szymku{\'{c}} S, Grzybowski BA.
  2020  Synthetic connectivity, emergence, and self-regeneration in the network
  of prebiotic chemistry. {\em Science} \textbf{369}.

\bibitem{Subramanian2020}
Subramanian H, Brown J, Gatenby R. 2020  Prebiotic competition and evolution in
  self-replicating polynucleotides can explain the properties of {DNA}/{RNA} in
  modern living systems. {\em {BMC} Evolutionary Biology} \textbf{20}.

\bibitem{horowitz2017spontaneous}
Horowitz JM, England JL. 2017  Spontaneous fine-tuning to environment in
  many-species chemical reaction networks. {\em Proceedings of the National
  Academy of Sciences} \textbf{114}, 7565--7570.

\bibitem{KAUFFMAN1969}
Kauffman S. 1969  Homeostasis and Differentiation in Random Genetic Control
  Networks. {\em Nature} \textbf{224}, 177--178.

\bibitem{marquez2021evolution}
M{\'a}rquez-Zacar{\'\i}as P, Pineau RM, Gomez M, Veliz-Cuba A, Murrugarra D,
  Ratcliff WC, Niklas KJ. 2021  Evolution of cellular differentiation: from
  hypotheses to models. {\em Trends in Ecology \& Evolution} \textbf{36},
  49--60.

\bibitem{Hutchison2016}
Hutchison CA, Chuang RY, Noskov VN, Assad-Garcia N, Deerinck TJ, Ellisman MH,
  Gill J, Kannan K, Karas BJ, Ma L, Pelletier JF, Qi ZQ, Richter RA,
  Strychalski EA, Sun L, Suzuki Y, Tsvetanova B, Wise KS, Smith HO, Glass JI,
  Merryman C, Gibson DG, Venter JC. 2016  Design and synthesis of a minimal
  bacterial genome. {\em Science} \textbf{351}.

\bibitem{Sol2007}
Sol{\'{e}} RV, Munteanu A, Rodriguez-Caso C, Mac{\'{\i}}a J. 2007  Synthetic
  protocell biology: from reproduction to computation. {\em Philosophical
  Transactions of the Royal Society B: Biological Sciences} \textbf{362},
  1727--1739.

\bibitem{Sol2018}
Sol{\'{e}} R, Oll{\'{e}}-Vila A, Vidiella B, Duran-Nebreda S, Conde-Pueyo N.
  2018  The road to synthetic multicellularity. {\em Current Opinion in Systems
  Biology} \textbf{7}, 60--67.

\bibitem{Kriegman2020}
Kriegman S, Blackiston D, Levin M, Bongard J. 2020  A scalable pipeline for
  designing reconfigurable organisms. {\em Proceedings of the National Academy
  of Sciences} \textbf{117}, 1853--1859.

\bibitem{Levin2020}
Levin M, Bongard J, Lunshof JE. 2020  Applications and ethics of
  computer-designed organisms. {\em Nature Reviews Molecular Cell Biology}
  \textbf{21}, 655--656.

\bibitem{Blackiston2021}
Blackiston D, Lederer E, Kriegman S, Garnier S, Bongard J, Levin M. 2021  A
  cellular platform for the development of synthetic living machines. {\em
  Science Robotics} \textbf{6}.

\bibitem{Wilke2001}
Wilke CO, Wang JL, Ofria C, Lenski RE, Adami C. 2001  Evolution of digital
  organisms at high mutation rates leads to survival of the flattest. {\em
  Nature} \textbf{412}, 331--333.

\bibitem{Lenski2003}
Lenski RE, Ofria C, Pennock RT, Adami C. 2003  The evolutionary origin of
  complex features. {\em Nature} \textbf{423}, 139--144.

\bibitem{tero2010rules}
Tero A, Takagi S, Saigusa T, Ito K, Bebber DP, Fricker MD, Yumiki K, Kobayashi
  R, Nakagaki T. 2010  Rules for biologically inspired adaptive network design.
  {\em Science} \textbf{327}, 439--442.

\bibitem{hopfield1982neural}
Hopfield JJ. 1982  Neural networks and physical systems with emergent
  collective computational abilities. {\em Proceedings of the National Academy
  of Sciences} \textbf{79}, 2554--2558.

\bibitem{OzkanAydin2021}
Ozkan-Aydin Y, Goldman DI, Bhamla MS. 2021  Collective dynamics in entangled
  worm and robot blobs. {\em Proceedings of the National Academy of Sciences}
  \textbf{118}, e2010542118.

\bibitem{chialvo1995swarms}
Chialvo DR, Millonas MM. 1995  How swarms build cognitive maps. In {\em The
  biology and technology of intelligent autonomous agents} pp. 439--450.
  Springer.

\bibitem{peleg2018collective}
Peleg O, Peters JM, Salcedo MK, Mahadevan L. 2018  Collective mechanical
  adaptation of honeybee swarms. {\em Nature Physics} \textbf{14}, 1193--1198.

\bibitem{Couzin2005}
Couzin ID, Krause J, Franks NR, Levin SA. 2005  Effective leadership and
  decision-making in animal groups on the move. {\em Nature} \textbf{433},
  513--516.

\bibitem{Sridhar2021}
Sridhar VH, Li L, Gorbonos D, Nagy M, Schell BR, Sorochkin T, Gov NS, Couzin
  ID. 2021  The geometry of decision-making in individuals and collectives.
  {\em Proceedings of the National Academy of Sciences} \textbf{118},
  e2102157118.

\bibitem{Couzin2009}
Couzin ID. 2009  Collective cognition in animal groups. {\em Trends in
  Cognitive Sciences} \textbf{13}, 36--43.

\bibitem{Hamilton1971}
Hamilton W. 1971  Geometry for the selfish herd. {\em Journal of Theoretical
  Biology} \textbf{31}, 295--311.

\bibitem{Caraco1980}
Caraco T, Martindale S, Pulliam HR. 1980  Avian flocking in the presence of a
  predator. {\em Nature} \textbf{285}, 400--401.

\bibitem{Clark1984}
Clark CW, Mangel M. 1984  Foraging and Flocking Strategies: Information in an
  Uncertain Environment. {\em The American Naturalist} \textbf{123}, 626--641.

\bibitem{Ryer1991}
Ryer CH, Olla BL. 1991  Information transfer and the facilitation and
  inhibition of feeding in a schooling fish. {\em Environmental Biology of
  Fishes} \textbf{30}, 317--323.

\bibitem{Katz2011}
Katz Y, Tunstrom K, Ioannou CC, Huepe C, Couzin ID. 2011  Inferring the
  structure and dynamics of interactions in schooling fish. {\em Proceedings of
  the National Academy of Sciences} \textbf{108}, 18720--18725.

\bibitem{Tunstrm2013}
Tunstr{\o}m K, Katz Y, Ioannou CC, Huepe C, Lutz MJ, Couzin ID. 2013
  Collective States, Multistability and Transitional Behavior in Schooling
  Fish. {\em {PLoS} Computational Biology} \textbf{9}, e1002915.

\bibitem{Bialek2012}
Bialek W, Cavagna A, Giardina I, Mora T, Silvestri E, Viale M, Walczak AM. 2012
   Statistical mechanics for natural flocks of birds. {\em Proceedings of the
  National Academy of Sciences} \textbf{109}, 4786--4791.

\bibitem{vicsek2012collective}
Vicsek T, Zafeiris A. 2012  Collective motion. {\em Physics Reports}
  \textbf{517}, 71--140.

\bibitem{Cavagna2013}
Cavagna A, Giardina I, Ginelli F. 2013  Boundary Information Inflow Enhances
  Correlation in Flocking. {\em Physical Review Letters} \textbf{110}.

\bibitem{Mora2016}
Mora T, Walczak AM, Castello LD, Ginelli F, Melillo S, Parisi L, Viale M,
  Cavagna A, Giardina I. 2016  Local equilibrium in bird flocks. {\em Nature
  Physics} \textbf{12}, 1153--1157.

\bibitem{Cavagna2018}
Cavagna A, Giardina I, Grigera TS. 2018  The physics of flocking: Correlation
  as a compass from experiments to theory. {\em Physics Reports} \textbf{728},
  1--62.

\bibitem{ball2004critical}
Ball P. 2004 {\em Critical mass: How one thing leads to another}.
Macmillan.

\bibitem{schelling1971dynamic}
Schelling TC. 1971  Dynamic models of segregation. {\em Journal of Mathematical
  Sociology} \textbf{1}, 143--186.

\bibitem{baronchelli2018emergence}
Baronchelli A. 2018  The emergence of consensus: a primer. {\em Royal Society
  Open Science} \textbf{5}, 172189.

\bibitem{helbing2001traffic}
Helbing D. 2001  Traffic and related self-driven many-particle systems. {\em
  Reviews of Modern Physics} \textbf{73}, 1067.

\bibitem{bettencourt2007growth}
Bettencourt LM, Lobo J, Helbing D, K{\"u}hnert C, West GB. 2007  Growth,
  innovation, scaling, and the pace of life in cities. {\em Proceedings of the
  National Academy of Sciences} \textbf{104}, 7301--7306.

\bibitem{bettencourt2013origins}
Bettencourt LM. 2013  The origins of scaling in cities. {\em Science}
  \textbf{340}, 1438--1441.

\bibitem{barthelemy2016structure}
Barthelemy M. 2016 {\em The structure and dynamics of cities}.
Cambridge University Press.

\bibitem{barthelemy2019statistical}
Barthelemy M. 2019  The statistical physics of cities. {\em Nature Reviews
  Physics} \textbf{1}, 406--415.

\bibitem{verbavatz2020growth}
Verbavatz V, Barthelemy M. 2020  The growth equation of cities. {\em Nature}
  \textbf{587}, 397--401.

\bibitem{boccaletti2006complex}
Boccaletti S, Latora V, Moreno Y, Chavez M, Hwang DU. 2006  Complex networks:
  Structure and dynamics. {\em Physics Reports} \textbf{424}, 175--308.

\bibitem{de2013mathematical}
De~Domenico M, Sol{\'e}-Ribalta A, Cozzo E, Kivel{\"a} M, Moreno Y, Porter MA,
  G{\'o}mez S, Arenas A. 2013  Mathematical formulation of multilayer networks.
  {\em Physical Review X} \textbf{3}, 041022.

\bibitem{artime2022multilayer}
Artime O, Benigni B, Bertagnolli G, d'Andrea V, Gallotti R, Ghavasieh A,
  Raimondo S, De~Domenico M. 2022 {\em Multilayer Network Science}.
Cambridge University Press.

\bibitem{suweis2013emergence}
Suweis S, Simini F, Banavar JR, Maritan A. 2013  Emergence of structural and
  dynamical properties of ecological mutualistic networks. {\em Nature}
  \textbf{500}, 449--452.

\bibitem{barabasi1999emergence}
Barab{\'a}si AL, Albert R. 1999  Emergence of scaling in random networks. {\em
  Science} \textbf{286}, 509--512.

\bibitem{d2019explosive}
D'Souza RM, G{\'o}mez-Gardenes J, Nagler J, Arenas A. 2019  Explosive phenomena
  in complex networks. {\em Advances in Physics} \textbf{68}, 123--223.

\bibitem{Simon1962}
Simon HA. 1962  The Architecture of Complexity. {\em Proceedings of the
  American Philosophical Society} \textbf{106}.

\bibitem{Simon1977}
Simon HA. 1977  The Organization of Complex Systems. In {\em Models of
  Discovery} pp. 245--261. Springer Netherlands.

\bibitem{newman2012communities}
Newman ME. 2012  Communities, modules and large-scale structure in networks.
  {\em Nature Physics} \textbf{8}, 25--31.

\bibitem{fortunato2016community}
Fortunato S, Hric D. 2016  Community detection in networks: A user guide. {\em
  Physics Reports} \textbf{659}, 1--44.

\bibitem{Peixoto2019}
Peixoto T. 2019  Bayesian Stochastic Blockmodeling. In {\em Advances in Network
  Clustering and Blockmodeling} pp. 289--332. Wiley.

\bibitem{ravasz2002hierarchical}
Ravasz E, Somera AL, Mongru DA, Oltvai ZN, Barab{\'a}si AL. 2002  Hierarchical
  organization of modularity in metabolic networks. {\em Science} \textbf{297},
  1551--1555.

\bibitem{ravasz2003hierarchical}
Ravasz E, Barab{\'a}si AL. 2003  Hierarchical organization in complex networks.
  {\em Physical Review E} \textbf{67}, 026112.

\bibitem{han2004evidence}
Han JDJ, Bertin N, Hao T, Goldberg DS, Berriz GF, Zhang LV, Dupuy D, Walhout
  AJ, Cusick ME, Roth FP et~al.. 2004  Evidence for dynamically organized
  modularity in the yeast protein--protein interaction network. {\em Nature}
  \textbf{430}, 88--93.

\bibitem{guimera2005functional}
Guimera R, Amaral LAN. 2005  Functional cartography of complex metabolic
  networks. {\em Nature} \textbf{433}, 895--900.

\bibitem{colizza2006detecting}
Colizza V, Flammini A, Serrano MA, Vespignani A. 2006  Detecting rich-club
  ordering in complex networks. {\em Nature Physics} \textbf{2}, 110--115.

\bibitem{hintze2008evolution}
Hintze A, Adami C. 2008  Evolution of complex modular biological networks. {\em
  PLoS Computational Biology} \textbf{4}, e23.

\bibitem{taylor2009dynamic}
Taylor IW, Linding R, Warde-Farley D, Liu Y, Pesquita C, Faria D, Bull S,
  Pawson T, Morris Q, Wrana JL. 2009  Dynamic modularity in protein interaction
  networks predicts breast cancer outcome. {\em Nature Biotechnology}
  \textbf{27}, 199--204.

\bibitem{bullmore2012economy}
Bullmore E, Sporns O. 2012  The economy of brain network organization. {\em
  Nature Reviews Neuroscience} \textbf{13}, 336--349.

\bibitem{ghavasieh2020statistical}
Ghavasieh A, Nicolini C, De~Domenico M. 2020  Statistical physics of complex
  information dynamics. {\em Physical Review E} \textbf{102}, 052304.

\bibitem{bertagnolli2021quantifying}
Bertagnolli G, Gallotti R, De~Domenico M. 2021  Quantifying efficient
  information exchange in real network flows. {\em Communications Physics}
  \textbf{4}, 1--10.

\bibitem{boguna2021network}
Boguna M, Bonamassa I, De~Domenico M, Havlin S, Krioukov D, Serrano M{\'A}.
  2021  Network geometry. {\em Nature Reviews Physics} \textbf{3}, 114--135.

\bibitem{gao2012networks}
Gao J, Buldyrev SV, Stanley HE, Havlin S. 2012  Networks formed from
  interdependent networks. {\em Nature Physics} \textbf{8}, 40--48.

\bibitem{mucha2010community}
Mucha PJ, Richardson T, Macon K, Porter MA, Onnela JP. 2010  Community
  structure in time-dependent, multiscale, and multiplex networks. {\em
  Science} \textbf{328}, 876--878.

\bibitem{boccaletti2014structure}
Boccaletti S, Bianconi G, Criado R, Del~Genio CI, G{\'o}mez-Gardenes J, Romance
  M, Sendina-Nadal I, Wang Z, Zanin M. 2014  The structure and dynamics of
  multilayer networks. {\em Physics Reports} \textbf{544}, 1--122.

\bibitem{kivela2014multilayer}
Kivel{\"a} M, Arenas A, Barthelemy M, Gleeson JP, Moreno Y, Porter MA. 2014
  Multilayer networks. {\em Journal of Complex Networks} \textbf{2}, 203--271.

\bibitem{de2016physics}
De~Domenico M, Granell C, Porter MA, Arenas A. 2016  The physics of spreading
  processes in multilayer networks. {\em Nature Physics} \textbf{12}, 901--906.

\bibitem{burda2009localization}
Burda Z, Duda J, Luck JM, Waclaw B. 2009  Localization of the maximal entropy
  random walk. {\em Physical Review Letters} \textbf{102}, 160602.

\bibitem{kuramoto1975international}
Kuramoto Y. 1975  International symposium on mathematical problems in
  theoretical physics. {\em Lecture notes in Physics} \textbf{30}, 420.

\bibitem{acebron2005kuramoto}
Acebr{\'o}n JA, Bonilla LL, Vicente CJP, Ritort F, Spigler R. 2005  The
  Kuramoto model: A simple paradigm for synchronization phenomena. {\em Reviews
  of Modern Physics} \textbf{77}, 137.

\bibitem{arenas2008synchronization}
Arenas A, D{\'\i}az-Guilera A, Kurths J, Moreno Y, Zhou C. 2008
  Synchronization in complex networks. {\em Physics Reports} \textbf{469},
  93--153.

\bibitem{rodrigues2016kuramoto}
Rodrigues FA, Peron TKD, Ji P, Kurths J. 2016  The Kuramoto model in complex
  networks. {\em Physics Reports} \textbf{610}, 1--98.

\bibitem{gomez2011explosive}
G{\'o}mez-Gardenes J, G{\'o}mez S, Arenas A, Moreno Y. 2011  Explosive
  synchronization transitions in scale-free networks. {\em Physical Review
  Letters} \textbf{106}, 128701.

\bibitem{o2017oscillators}
O'Keeffe KP, Hong H, Strogatz SH. 2017  Oscillators that sync and swarm. {\em
  Nature Communications} \textbf{8}, 1--13.

\bibitem{ghosh2022synchronized}
Ghosh D, Frasca M, Rizzo A, Majhi S, Rakshit S, Alfaro-Bittner K, Boccaletti S.
  2022  The synchronized dynamics of time-varying networks. {\em Physics
  Reports} \textbf{949}, 1--63.

\bibitem{granell2013dynamical}
Granell C, G{\'o}mez S, Arenas A. 2013  Dynamical interplay between awareness
  and epidemic spreading in multiplex networks. {\em Physical Review Letters}
  \textbf{111}, 128701.

\bibitem{sanz2014dynamics}
Sanz J, Xia CY, Meloni S, Moreno Y. 2014  Dynamics of interacting diseases.
  {\em Physical Review X} \textbf{4}, 041005.

\bibitem{castioni2021critical}
Castioni P, Gallotti R, De~Domenico M. 2021  Critical behavior in
  interdependent spatial spreading processes with distinct characteristic time
  scales. {\em Communications Physics} \textbf{4}, 1--10.

\bibitem{barrat2008dynamical}
Barrat A, Barthelemy M, Vespignani A. 2008 {\em Dynamical processes on complex
  networks}.
Cambridge University Press.

\bibitem{li2021percolation}
Li M, Liu RR, L{\"u} L, Hu MB, Xu S, Zhang YC. 2021  Percolation on complex
  networks: Theory and application. {\em Physics Reports}.

\bibitem{albert2000error}
Albert R, Jeong H, Barab{\'a}si AL. 2000  Error and attack tolerance of complex
  networks. {\em Nature} \textbf{406}, 378--382.

\bibitem{holme2002attack}
Holme P, Kim BJ, Yoon CN, Han SK. 2002  Attack vulnerability of complex
  networks. {\em Physical Review E} \textbf{65}, 056109.

\bibitem{grassia2021machine}
Grassia M, De~Domenico M, Mangioni G. 2021  Machine learning dismantling and
  early-warning signals of disintegration in complex systems. {\em Nature
  Communications} \textbf{12}.

\bibitem{artime2021percolation}
Artime O, De~Domenico M. 2021  Percolation on feature-enriched interconnected
  systems. {\em Nature Communications} \textbf{12}, 1--12.

\bibitem{goltsev2006k}
Goltsev AV, Dorogovtsev SN, Mendes JFF. 2006  $k$-core (bootstrap) percolation
  on complex networks: Critical phenomena and nonlocal effects. {\em Physical
  Review E} \textbf{73}, 056101.

\bibitem{baxter2010bootstrap}
Baxter GJ, Dorogovtsev SN, Goltsev AV, Mendes JF. 2010  Bootstrap percolation
  on complex networks. {\em Physical Review E} \textbf{82}, 011103.

\bibitem{achlioptas2009explosive}
Achlioptas D, D'Souza RM, Spencer J. 2009  Explosive percolation in random
  networks. {\em Science} \textbf{323}, 1453--1455.

\bibitem{son2012percolation}
Son SW, Bizhani G, Christensen C, Grassberger P, Paczuski M. 2012  Percolation
  theory on interdependent networks based on epidemic spreading. {\em
  Europhysics Letters} \textbf{97}, 16006.

\bibitem{radicchi2015percolation}
Radicchi F. 2015  Percolation in real interdependent networks. {\em Nature
  Physics} \textbf{11}, 597--602.

\bibitem{watts2002simple}
Watts DJ. 2002  A simple model of global cascades on random networks. {\em
  Proceedings of the National Academy of Sciences} \textbf{99}, 5766--5771.

\bibitem{motter2002cascade}
Motter AE, Lai YC. 2002  Cascade-based attacks on complex networks. {\em
  Physical Review E} \textbf{66}, 065102.

\bibitem{brummitt2012suppressing}
Brummitt CD, D’Souza RM, Leicht EA. 2012  Suppressing cascades of load in
  interdependent networks. {\em Proceedings of the National Academy of
  Sciences} \textbf{109}, E680--E689.

\bibitem{yu2016system}
Yu Y, Xiao G, Zhou J, Wang Y, Wang Z, Kurths J, Schellnhuber HJ. 2016  System
  crash as dynamics of complex networks. {\em Proceedings of the National
  Academy of Sciences} \textbf{113}, 11726--11731.

\bibitem{huang2013cascading}
Huang X, Vodenska I, Havlin S, Stanley HE. 2013  Cascading failures in
  bi-partite graphs: model for systemic risk propagation. {\em Scientific
  Reports} \textbf{3}, 1--9.

\bibitem{shekhtman2016recent}
Shekhtman LM, Danziger MM, Havlin S. 2016  Recent advances on failure and
  recovery in networks of networks. {\em Chaos, Solitons \& Fractals}
  \textbf{90}, 28--36.

\bibitem{buldyrev2010catastrophic}
Buldyrev SV, Parshani R, Paul G, Stanley HE, Havlin S. 2010  Catastrophic
  cascade of failures in interdependent networks. {\em Nature} \textbf{464},
  1025--1028.

\bibitem{de2014navigability}
De~Domenico M, Sol{\'e}-Ribalta A, G{\'o}mez S, Arenas A. 2014  Navigability of
  interconnected networks under random failures. {\em Proceedings of the
  National Academy of Sciences} \textbf{111}, 8351--8356.

\bibitem{chalmers2006}
Chalmers D. 2006  Strong and Weak Emergence. In Davies P, Clayton P, editors,
  {\em The Re-Emergence of Emergence: The Emergentist Hypothesis From Science
  to Religion}. Oxford University Press.

\bibitem{bedau1997weak}
Bedau MA. 1997  Weak emergence. {\em Philosophical perspectives} \textbf{11},
  375--399.

\bibitem{gardner1970fantastic}
Gardner M. 1970  The Fantastic Combinations of John Conway's New Solitaire Game
  ``Life''. {\em Scientific American} \textbf{223}, 20--123.

\bibitem{ney2022entanglement}
Ney PM, Notarnicola S, Montangero S, Morigi G. 2022  Entanglement in the
  quantum Game of Life. {\em Physical Review A} \textbf{105}, 012416.

\bibitem{abrahao2022emergence}
Abrah\~{a}o FS, Zenil H. 2022  Emergence and algorithmic information dynamics
  of systems and observers. {\em Philosophical Transactions of the Royal
  Society A: Mathematical, Physical and Engineering Sciences} \textbf{380},
  20200429.

\bibitem{rosas2022greater}
Mediano PAM, Rosas FE, Luppi AI, Jensen HJ, Seth AK, Barrett AB, Carhart-Harris
  RL, Bor D. 2022  Greater than the parts: A review of the information
  decomposition approach to causal emergence. {\em Philosophical Transactions
  of the Royal Society A: Mathematical, Physical and Engineering Sciences}
  \textbf{380}, 20210246.

\bibitem{varley2022emergence}
Varley TF, Hoel E. 2022  Emergence as the conversion of information: A unifying
  theory. {\em Philosophical Transactions of the Royal Society A: Mathematical,
  Physical and Engineering Sciences} \textbf{380}, 20210150.

\bibitem{adami2022emergence}
Adami C, Nitash CG. 2022  Emergence of functional information from multivariate
  correlations. {\em Philosophical Transactions of the Royal Society A:
  Mathematical, Physical and Engineering Sciences} \textbf{380}, 20210250.

\bibitem{wolpert2022strengthened}
Wolpert DH. 2022  Strengthened second law for multi-dimensional systems coupled
  to multiple thermodynamic reservoirs. {\em Philosophical Transactions of the
  Royal Society A: Mathematical, Physical and Engineering Sciences}
  \textbf{380}, 20200428.

\bibitem{sokolov2022emergent}
Sokolov B, Rossi MAC, Garc\'{i}a-P\'{e}rez G, Maniscalco S. 2022  Emergent
  entanglement structures and self-similarity in quantum spin chains. {\em
  Philosophical Transactions of the Royal Society A: Mathematical, Physical and
  Engineering Sciences} \textbf{380}, 20200421.

\bibitem{nader2022protometabolism}
Nader S, Sebastianelli L, Mansy SS. 2022  Protometabolism as out-of-equilibrium
  chemistry. {\em Philosophical Transactions of the Royal Society A:
  Mathematical, Physical and Engineering Sciences} \textbf{380}, 20200423.

\bibitem{xavier2022small}
Xavier JC, Kauffman S. 2022  Small-molecule autocatalytic networks are
  universal metabolic fossils. {\em Philosophical Transactions of the Royal
  Society A: Mathematical, Physical and Engineering Sciences} \textbf{380},
  20210244.

\bibitem{manrubia2022simple}
Manrubia S. 2022  The simple emergence of complex molecular function. {\em
  Philosophical Transactions of the Royal Society A: Mathematical, Physical and
  Engineering Sciences} \textbf{380}, 20200422.

\bibitem{vohryzek2022understanding}
Vohryzek J, Cabral J, Vuust P, Deco G, Kringelbach ML. 2022  Understanding
  brain states across spacetime informed by whole-brain modelling. {\em
  Philosophical Transactions of the Royal Society A: Mathematical, Physical and
  Engineering Sciences} \textbf{380}, 20200247.

\bibitem{buendia2022broad}
Buend\'{i}a V, Villegas P, Burioni R, Mu\~{n}oz MA. 2022  The broad edge of
  synchronisation: Griffiths effects and collective phenomena in brain
  networks. {\em Philosophical Transactions of the Royal Society A:
  Mathematical, Physical and Engineering Sciences} \textbf{380}, 20200424.

\bibitem{saeedian2022effect}
Saeedian M, Pigani E, Maritan A, Suweis S, Azaele S. 2022  Effect of delay on
  the emergent stability patterns in Lotka-Volterra ecological dynamics. {\em
  Philosophical Transactions of the Royal Society A: Mathematical, Physical and
  Engineering Sciences} \textbf{380}, 20210245.

\bibitem{peters2022ergodicity}
Peters O, Adamou A. 2022  The ergodicity solution of the cooperation puzzle.
  {\em Philosophical Transactions of the Royal Society A: Mathematical,
  Physical and Engineering Sciences} \textbf{380}, 20210425.

\bibitem{andrighetto2022research}
Andrighetto G, Vriens E. 2022  A research agenda for the study of social norm
  change. {\em Philosophical Transactions of the Royal Society A: Mathematical,
  Physical and Engineering Sciences} \textbf{380}, 20210411.

\bibitem{brinkmann2022hybrid}
Brinkmann L, Gezerli D, Kleist Kv, M\"{u}ller TF, Rahwan I, Pescetelli N. 2022
  Hybrid social learning in human-algorithm cultural transmission. {\em
  Philosophical Transactions of the Royal Society A: Mathematical, Physical and
  Engineering Sciences} \textbf{380}, 20200426.

\bibitem{khanjanianpak2022emergence}
Khanjanianpak M, Azimi-Tafreshi N, Arenas A, G\'{o}mez-Garde\~{n}es J. 2022
  Emergence of protective behaviour under different risk perceptions to disease
  spreading. {\em Philosophical Transactions of the Royal Society A:
  Mathematical, Physical and Engineering Sciences} \textbf{380}, 20200412.

\bibitem{Ramalho2021}
Ramalho T, Kremser S, Wu H, Gerland U. 2021  Programmable pattern formation in
  cellular systems with local signaling. {\em Communications Physics}
  \textbf{4}.

\bibitem{Rubenstein2014}
Rubenstein M, Cornejo A, Nagpal R. 2014  Programmable self-assembly in a
  thousand-robot swarm. {\em Science} \textbf{345}, 795--799.

\bibitem{Li2019}
Li S, Batra R, Brown D, Chang HD, Ranganathan N, Hoberman C, Rus D, Lipson H.
  2019  Particle robotics based on statistical mechanics of loosely
  coupled~components. {\em Nature} \textbf{567}, 361--365.

\bibitem{brambilla2013swarm}
Brambilla M, Ferrante E, Birattari M, Dorigo M. 2013  Swarm robotics: a review
  from the swarm engineering perspective. {\em Swarm Intelligence} \textbf{7},
  1--41.

\bibitem{Vsrhelyi2018}
V{\'{a}}s{\'{a}}rhelyi G, Vir{\'{a}}gh C, Somorjai G, Nepusz T, Eiben AE,
  Vicsek T. 2018  Optimized flocking of autonomous drones in confined
  environments. {\em Science Robotics} \textbf{3}.

\bibitem{Savoie2019}
Savoie W, Berrueta TA, Jackson Z, Pervan A, Warkentin R, Li S, Murphey TD,
  Wiesenfeld K, Goldman DI. 2019  A robot made of robots: Emergent transport
  and control of a smarticle ensemble. {\em Science Robotics} \textbf{4}.

\bibitem{haeger2015collective}
Haeger A, Wolf K, Zegers MM, Friedl P. 2015  Collective cell migration:
  guidance principles and hierarchies. {\em Trends in Cell Biology}
  \textbf{25}, 556--566.

\bibitem{Aguilar2018}
Aguilar J, Monaenkova D, Linevich V, Savoie W, Dutta B, Kuan HS, Betterton MD,
  Goodisman MAD, Goldman DI. 2018  Collective clog control: Optimizing traffic
  flow in confined biological and robophysical excavation. {\em Science}
  \textbf{361}, 672--677.

\bibitem{WarnatHerresthal2021}
Warnat-Herresthal S, Schultze H, Shastry K, {et al}. 2021  Swarm Learning for
  decentralized and confidential clinical machine learning. {\em Nature}
  \textbf{594}, 265--270.

\end{thebibliography}

\end{document}